\documentclass[preprint]{aastex} 

\shorttitle{Secular Interactions} 
\shortauthors{F. C. Adams} 
 
\newcommand{\be}{\begin{equation}}
\newcommand{\ee}{\end{equation}}
\newcommand{\ebar}{{\langle e \rangle}} 
\newcommand{\fbar}{{\langle F \rangle}} 
\newcommand{\albar}{{\bar \alpha}} 
\newcommand{\eigenv}{{ \Lambda }}

\newcommand{\taugr}{{ \tau_{\rm gr} }} 
\newcommand\sgn{ {\rm sign}} 
 
\newcommand\mel{ {\hat m} } 
\def\lta{\,\raise 0.3 ex\hbox{$ < $}\kern -0.75 em
 \lower 0.7 ex\hbox{$\sim$}\,}
\def\gta{\,\raise 0.3 ex\hbox{$ > $}\kern -0.75 em
 \lower 0.7 ex\hbox{$\sim$}\,} 

\begin{document}
 
\title{Effects of Secular Interactions 
in Extrasolar Planetary Systems} 
 
\author{Fred C. Adams$^{1,2}$ and Gregory Laughlin$^3$} 

\affil{$^1$Michigan Center for Theoretical Physics \\
Physics Department, University of Michigan, Ann Arbor, MI 48109} 
 
\affil{$^2$Astronomy Department, University of Michigan, Ann Arbor, MI 48109}

\affil{$^3$Lick Observatory, University of California, Santa Cruz, CA 95064} 
 
\begin{abstract} 

This paper studies the effects of dynamical interactions among the
planets in observed extrasolar planetary systems, including
hypothetical additional bodies, with a focus on secular perturbations.
These interactions cause the eccentricities of the planets to explore
a distribution of values over time scales that are long compared to
observational time baselines, but short compared to the age of the
systems.  The same formalism determines the eccentricity forcing of
hypothetical test bodies (terrestrial planets) in these systems and we
find which systems allow for potentially habitable planets.  Such
planets would be driven to nonzero orbital eccentricity and we derive
the distribution of stellar flux experienced by the planets over the
course of their orbits.  The general relativistic corrections to
secular interaction theory are included in the analysis and such
effects are important in systems with close planets ($\sim$4 day
orbits). Some extrasolar planetary systems (e.g., Upsilon Andromedae)
can be used as a test of general relativity, whereas in other systems,
general relativity can be used to constrain the system parameters
(e.g., $\sin i \gta 0.93$ for HD160691). For the case of hot Jupiters,
we discuss how the absence of observed eccentricity implies the
absence of companion planets.

\end{abstract}


\section{Introduction} 
 
The number of discovered planets in other systems is rapidly growing
and the observational sample shows an astonishing diversity of
architectures (beginning with Mayor \& Queloz 1995; Marcy \& Butler
1996). One way to characterize these planetary systems is in terms of
their orbital elements, where the semi-major axis $a$ and orbital
eccentricity $e$ are most often used. These variables are equivalent
to specifying the energy and angular momentum of the orbit. Previous
expectations, informed by the structure of our Solar System, predicted
planetary orbits with larger values of $a$ and smaller values of $e$
than those found in the current observational sample. A major
theoretical effort is now being put forth to provide an explanation of
the observed distributions of orbital elements, e.g., the $a-e$ plane.
In multiple planet systems, however, it can be important to keep in
mind that the orbital eccentricities can vary dramatically through
secular interactions on time scales that are long compared to
observational baselines but short compared to the system ages.
Instead of being described by a single value of eccentricity, the
orbits of planets in multiple planet systems should generally be
characterized by a complete distribution of eccentricity values (see
also Takeda \& Rasio 2005).

Secular interactions, and the apparent lack of evidence for such
interactions, can be used to place constraints on observed planetary
systems. For the known multiple planet systems, secular interactions
place strong constraints on the possibility of additional as-yet
undetected terrestrial planets. For putative single planet systems
with zero or low eccentricity, the inferred absence of secular
interactions (which tend to prevent small eccentricity values $e \sim
0$) implies that any additional planets in the system must have small
masses, small eccentricities, and/or large semi-major axes; we
quantify these constraints for observed systems. For multiple planet
systems with hot Jupiters (close planetary companions, e.g., in 4 day
orbits), secular perturbations tend to increase the eccentricity of
the inner planet, which also experiences tidal stressing forces from
the star. This interplay leads to continual energy dissipation in the
planet. 

This paper follows a large body of previous work. Secular interactions
have been studied for centuries, primarily in the context of our Solar
System (Brower \& van Woerkom 1950; Laskar 1988; and many others).
With the discovery of extrasolar multiple planet systems (starting
with Upsilon Andromedae; Butler et al. 1999), secular perturbation
theory can be applied to an ever growing collection of planetary
systems (e.g., Wu \& Goldreich 2002).  General schemes for studying
secular interactions have been developed (Mardling \& Lin 2002) and
used to study close (hypothetical) terrestrial planets (Mardling \&
Lin 2004). For hierarchical planetary systems, a higher order
(octopole) approximation scheme has been developed and applied to
HD168443 and HD12661 (Lee \& Peale 2003) and to triple star systems
(Ford et al. 2000ab); an alternate approximation scheme has been
developed and applied to the Upsilon Andromedae planetary system
(Michtchenko \& Malhotra 2004). The effects of disk potentials, and
the disappearance of the disk, have been studied (Nagasawa et al.
2003) with the goal of explaining observed orbital eccentricities (see
also Lubow \& Ogilvie 2001). The effects of the Kozai mechanism have
been explored for systems containing a binary companion (Takeda \&
Rasio 2005).

This paper builds upon the studies outlined above with the modest goal
of using secular interaction theory to extract additional information
from extant multiple planet systems.  The basic theory is outlined in
\S 2, which includes the leading order corrections for general
relativity (GR).  The results and applications are then presented in
\S 3.  In \S 3.1, we determine the distributions of eccentricity
sampled by extrasolar planets over the course of their secular cycles,
determine the secular time scales, and study the signature of secular
eccentricity variations in the $a-e$ diagram. We place constraints on
the possibility of additional terrestrial planets residing in
currently detected extrasolar planetary systems (\S 3.2) and place
constraints on possible additional giant planets in observed hot
Jupiter systems (\S 3.4).  We explore the effects of general
relativity in secular interactions and show that some systems can be
used to test relativity, while in other systems relativity can be used
to place constraints on system parameters (\S 3.3). We conclude, in \S
4, with a discussion and summary of results.

\section{Basic Theory of Secular Interactions} 

In this section we outline the basic theory of secular interactions as
applied to the planetary systems of interest. Since this topic has been
well studied over the past 250 years, the review is brief, and the
results are presented for the purpose of further development in later
sections.

To the second order in eccentricity and inclination angle, the
equations of motion for eccentricity $e_j$ and argument of periastron
$\varpi_j$ decouple from those of inclination angle and the ascending
node. Following standard convention (Murray \& Dermott 1999; hereafter
MD99), we work in terms of the variables defined by
\be
h_j \equiv e_j \sin \varpi_j \,  \qquad {\rm and} \qquad 
k_j \equiv e_j \cos \varpi_j \, , 
\ee 
where the subscript $j$ refers to the $jth$ planet in an $N$ planet 
system. The basic equations of motion for the theory can then be 
written in the form 
\be 
{d h_j \over dt} = {1 \over n_j a_j^2} 
{\partial {\cal R}_j \over \partial k_j} 
\qquad {\rm and} \qquad 
{d k_j \over dt} = - {1 \over n_j a_j^2} 
{\partial {\cal R}_j \over \partial h_j} \, , 
\ee 
where ${\cal R}_j$ the secular part of the disturbing function (see
below and MD99 for further detail), $n_j$ is the mean motion of the
$jth$ planet and $a_j$ is its corresponding semi-major axis. To
consistent order in this approximation, the relevant terms in the
disturbing function take the form 
\be 
{\cal R}_j = n_j a_j^2 \Bigl[ {1 \over 2} A_{jj} e_j^2 + 
\sum_{k \ne j} A_{jk} e_j e_k \cos (\varpi_j - \varpi_k) 
\Bigr] \, . 
\ee 
The physics of these interactions is thus encapsulated in the $N
\times N$ matrix $A_{ij}$, where the number $N$ of planets in the
system is usually $N$ = 2 or 3 for the systems observed to date.  
The matrix elements can be written in the form 
\be 
A_{jj} = n_j \Bigl[ {1 \over 4} \sum_{k \ne j} {m_k \over M_\ast + m_j} 
\alpha_{jk} \albar_{jk} b^{(1)}_{3/2}(\alpha_{jk}) \, + 
3 {G M_\ast \over c^2 a_j}  \Bigr] \, , 
\label{eq:diag} 
\ee 
and 
\be
A_{jk} = - n_j {1 \over 4} {m_k \over M_\ast + m_j} 
\alpha_{jk} \albar_{jk} b^{(2)}_{3/2}(\alpha_{jk}) \, .
\label{eq:offdiag} 
\ee
In the diagonal matrix elements (eq. [\ref{eq:diag}]), we have
included the leading order correction for general relativity ($c$ is
the speed of light).  Although these terms are small, $\mu \equiv
GM_\ast /(c^2 a_j) \ll 1$, such small corrections to the
eigenfrequencies can be important, especially when the system is near
resonance. The quantities $\alpha_{jk}$ are defined such that
$\alpha_{jk} = a_j/a_k$ ($a_k/a_j$) if $a_j < a_k$ ($a_k < a_j$). The
complementary quantities $\albar_{jk}$ are defined so that
$\albar_{jk} = a_j/a_k = \alpha_{jk}$ if $a_j < a_k$, but
$\albar_{jk}$ = 1 for $a_k < a_j$.  Finally, the quantities
$b^{(1)}_{3/2}(\alpha_{jk})$ and $b^{(2)}_{3/2}(\alpha_{jk})$ are
Laplace coefficients (MD99).

With the above definitions, the resulting solution takes the form 
\be 
h_j = \sum_i \eigenv_{ji} \sin (\lambda_i t + \beta_i)\, , \qquad 
k_j = \sum_i \eigenv_{ji} \cos (\lambda_i t + \beta_i)\, , 
\label{eq:solution} 
\ee
where the $\lambda_i$ are eigenvalues of the matrix $A_{ij}$ 
and the $\eigenv_{ji}$ are the corresponding eigenvectors. The phases 
$\beta_i$ and the normalization of the eigenvectors are determined 
by the initial conditions, i.e., the values of eccentricity $e_j$ 
and argument of periastron $\varpi_j$ for each planet at $t=0$.  

\section{Applications to Extrasolar Planetary Systems} 

In this section we use the formal development reviewed above to study
observed extrasolar planetary systems. First, we use the theory of
secular interactions to show the relationship between the observed
values of eccentricity and the underlying distribution of
eccentricities that characterize the systems (\S 3.1). Next we use
interactions, and their absence, to place new constraints on observed
multiple planet systems. We can find constraints on the possible
existence of additional small (terrestrial) planets in these systems
by requiring that any such planets must reside far from a secular
resonance (\S 3.2).  For a particular subset of systems, we can find
constraints on the viewing angle of the system, again by requiring
that the systems are far from resonance (\S 3.3); in this latter
application, the effects of general relativity are important and we
elucidate the possible GR signatures in other planetary systems. In
particular, some systems can be used as a test of general relativity.
Finally, we can place constraints on observed systems with hot
Jupiters and low (or zero) eccentricities, as typically found for such
systems. In order for the eccentricities to stay low, any additional
planets must be sufficiently distant (\S 3.4).

\subsection{Eccentricity Distributions and Secular Time Scales} 

As an application of secular interaction theory, we use the formalism
described above (see MD99 for further detail) to calculate the
variations in eccentricity in a sub-sample of observed extrasolar
planetary systems. First, the eigenvalues $\lambda_j$ for the multiple
planet systems were calculated and converted into time scales for a
collection of 19 observed multiple planet systems\footnote{The orbital
elements for these planets, along with predicted transit ephemerides
and other information are tabulated at www.transitsearch.org}.  The
results are listed in Table 1. Most of these multiple planet systems
have secular interaction time scales in the range $10^3 - 10^5$ yr.
These time scales are much longer than any possible observational
baseline (tens of years), but much shorter than the system lifetimes
(which are typically several Gyr). The shortest secular time scale
occurs for the GJ 876 system. Although the dynamics of this system are
dominated by the 2:1 resonance between planets ``c'' and ``b'', the  
secular interaction time $\tau_3$ = 4.4 yr gives an excellent estimate
of the time scale for dynamical interaction in the system. Indeed,
radial velocity fits to the system must take into account the
planet-planet interactions in order to obtain an acceptable fit (see,
e.g., Rivera et al. 2005 and the references within).

These secular interaction times are thus long enough that observations
can determine the eccentricity (and longitude of periastron) with high
accuracy at the present epoch. Over much longer time scales that are
not observationally accessible, however, the eccentricity (and
longitude of periastron) will vary according to the appropriate
secular cycles. As a result, attempts to explain the observed $a-e$
plane must take the possibility of secular variations into account.

The net effect of secular interactions on our observational
interpretation of these systems is that the measured eccentricity
values are a particular sampling of an underlying distribution.
Within the context of leading order secular theory, the distribution
of eccentricity is determined by the formalism of \S 2. For each of
the observed multiple planet systems considered in this paper, we have
calculated the expected time variations of eccentricity and longitude
of periastron according to secular theory. From this time series, we
have extracted the mean eccentricity $\ebar$, the variance $\sigma_e$
of the distribution, the minimum eccentricity value $e_{min}$, and the
maximum value $e_{max}$. The results are listed in Tables 2 and 3
along with the (currently) observed values of eccentricity. The
difference between the observed values and the mean eccentricities
averaged over many secular cycles can be substantial (more than a
factor of two). The width of the distribution can also be significant.
For the second planet of the HD160691 system, for example, the first
two moments of the eccentricity distribution imply $e = 0.49 \pm 0.19$. 

For the case of two planet systems, the formalism produces simple
analytic expressions for the parameters of the eccentricity
distribution. The distribution itself can be derived by taking the
solution of equation (\ref{eq:solution}) and solving for the
eccentricity as a function of time. Since time is distributed 
uniformly, the resulting expression can be solved for the
corresponding distribution of eccentricity, which can then be 
written in the form
\be 
{dP \over de} = \Bigl[ 1 - \bigl( {e^2 - \eigenv_{j1}^2 - \eigenv_{j2}^2 
\over 2 \eigenv_{j1} \eigenv_{j2}} \bigr)^2 \Bigr]^{-1/2} 
{e \over \pi \eigenv_{j1} \eigenv_{j2} } \, , 
\label{eq:dpde} 
\ee
where $\eigenv_{ji}$ are the eigenvectors. This form of the eccentricity 
distribution (for the $jth$ planet) is valid between the extremes given by 
\be 
e_{max}, e_{min} = \big| \eigenv_{j1} \pm \eigenv_{j2} \big| \, . 
\ee 
The mean value of the distribution can be evaluated from its 
definition $\ebar = \int e (dP/de) de$ and takes the form 
\be 
\ebar_j = (2/\pi) (\eigenv_{j1} + \eigenv_{j2}) 
E (\mel) \, , 
\ee 
where $E(\mel)$ is the elliptical integral of the second kind (Abramowitz
\& Stegun 1970) with parameter $\mel \equiv 4 \eigenv_{j1} \eigenv_{j2} / 
(\eigenv_{j1} + \eigenv_{j2})^2$.  Notice that the parameter $\mel$ can 
be negative and hence care must be taken in evaluating $E(\mel)$. The 
corresponding variance of the distribution is given by 
\be 
\sigma_{ej}^2 = \eigenv_{j1}^2 + \eigenv_{j2}^2 - {4 \over \pi^2} 
(\eigenv_{j1} + \eigenv_{j2})^2 \bigl[ E (\mel) \bigr]^2 \, . 
\label{eq:sigmae} 
\ee 
For two-planet systems, we have verified that these expressions for
the mean, extrema, and variance of the distribution are in good
agreement with those found via sampling of the secular solutions as
described above.

Figure \ref{fig:aeplane} illustrates the effect of this eccentricity
variation on the $a-e$ plane.  The bottom panel shows the locations of
the planets in the $a-e$ plane for our sub-sample of multiple planet
systems, where the observed eccentricities are taken at face
value. The top panel shows the same diagram with the observed
eccentricity values replaced by the ranges of eccentricities predicted
by secular theory. The relatively large (and previously unexpected)
values of eccentricity found in the observational sample to date has
led to an explosion of theoretical explanations (e.g., Ogilvie \&
Lubow 2003, Goldreich \& Sari 2003, Adams \& Laughlin 2003; Moorhead
\& Adams 2005).  Secular interactions change the target of these
investigations: Instead of explaining the observed distribution of
eccentricity values (where the observed $e$ are taken as given), a
complete theory must provide an explanation of the underlying
distribution of eccentricity from which the currently observed values
are sampled.

The secular theory used here is derived from a disturbing function
that is correct only to second order in eccentricity and inclination
angle and first order in the planet masses (MD99). In order to test
how well this version of secular theory describes the eccentricity
evolution of these systems, we have performed a set of numerical
integrations for a representative sub-sample of the extrasolar
planetary systems. The time evolution of eccentricity is shown in
Figures \ref{fig:numhd37124} -- \ref{fig:numhd12661} for three
representative systems in our sample (HD37124, HD168443, HD12661).
For HD37124 and HD168443 (Figures \ref{fig:numhd37124} and
\ref{fig:numhd168443}), the time evolution predicted by secular theory
is in good agreement with the numerical integrations, although not
exact.  For the HD12661 system, however, the eccentricity times
series produced by secular theory does not provide an accurate
description of that indicated by numerical integration. In addition,
the numerical integration results depend sensitively on the initial
conditions; Figure \ref{fig:numhd12661} shows the results of two sets
of starting parameters (both within the observational errors). In
spite of these complications, however, for all three of the systems,
the amplitude of eccentricity variation predicted by secular theory is
in good agreement with that predicted by direct numerical integration.
Specifically, the envelopes of the eccentricity variations are in
agreement. (Of course, an improved point-to-point time variation can
be obtained with a higher order approximation scheme -- see Lee \&
Peale 2003).  

\subsection{Possible Earth-like Planets in Extrasolar Planetary Systems} 

\subsubsection{Constraints from forced eccentricity oscillations} 

One of the themes of upcoming space missions (especially Terrestrial
Planet Finder, hereafter TPF) is to detect Earth-like planets, with an
emphasis on those in ``habitable'' Earth-like orbits. For solar mass
stars, the standard scenario (Kasting et al. 1993) suggests that such
orbits would have semi-major axes near $a = 1$ AU and relatively low
eccentricity (although the maximum eccentricity threshold is not well
determined -- see below). For the multiple planet systems discovered
to date, the detected planets are of order Jovian mass, much larger
than the mass of Earth. As a result, to first approximation, any
putative Earth-like planets present in these systems can be considered
as test particles. Within the framework of secular interactions
considered here, we can calculate the effects of extant planets on
hypothetical Earth-like planets (test bodies) that may reside in the
systems. According to secular theory (MD99, \S 2), the motion of test
particles can be characterized by a frequency $A$ of free oscillation,
which takes the form
\be 
A = n_0 \Bigl[ {1 \over 4} \sum_{j} {m_j \over M_\ast} 
\alpha_{0j} \albar_{0j} b^{(1)}_{3/2}(\alpha_{0j}) \, + 
3 {G M_\ast \over c^2 a_0}  \Bigr] \, ,  
\label{eq:afree} 
\ee 
where the subscript `0' refers to the orbit of the test particle and
where the rest of the symbols have the same meaning as in the previous
section. Notice that the free oscillation frequency $A$ is a function of
the test particle semi-major axis $a_0$.

By applying the formalism described above to extant systems with
multiple giant planets, we can explore the forced oscillations of any
additional small (e.g., terrestrial) planets. The results are shown in
Figures \ref{fig:reshd38529} -- \ref{fig:reshd12661} for a collection
of observed systems. This subset was chosen to include systems that
experience nontrivial secular interactions, but still might allow for
a habitable planet (HD38529, HD74156, HD168443). Other systems (not
shown) either have little secular interaction or already have a planet
near $a \sim 1$ AU (we consider HD12661 as an example of this latter
type of system). The bottom panels of these figures show the free
oscillation frequency $A$, plotted as a function of possible
semi-major axis $a_0$, for hypothetical test bodies in the given
extrasolar planetary systems.

A related quantity is the amplitude of forced eccentricity
oscillations; these amplitudes are plotted as a function of semi-major
axis in the top panels of Figures \ref{fig:reshd38529} --
\ref{fig:reshd12661}. Briefly, the eccentricity of a test particle at
any given time is the vector sum of the forced eccentricity and the
free eccentricity (see Fig. 7.2 of MD99).  The amplitude of forced
eccentricity oscillation is determined by the semi-major axis of the
test particle and the secular solution for the perturbing bodies; it
varies over secular time scales, as the perturbing planets interact
with each other. The free eccentricity is determined by the boundary
conditions and represents a more ``fundamental'' orbital parameter 
(for further detail see MD99).  The amplitudes shown in the figures
are calculated for the starting time $t=0$, i.e., with the current
values of the orbital elements.  Over time, the amplitude curve
oscillates and thereby gives an effective width to the curve.

For the systems HD38529, HD74156, HD168443, and HD37124 (Figs.
\ref{fig:reshd38529} -- \ref{fig:reshd37124}), the theory of secular
interactions predicts that Earth-like planets (small bodies with
semi-major axes near 1 AU) can reside in the planetary systems [both
$A$ and $e$(forced) are small]. However, the forced eccentricities of
the planets are appreciably larger than that of Earth, specifically in
the range $e$(forced) $\approx 0.10 - 0.30$. In addition, the outer
planets pose a threat to long term stability in these systems.
Extrapolation of the results of numerical integrations (David et
al. 2003) indicates that an Earth-like planet can remain stable over
the current 4.6 Gyr age of the Solar System for a Jupiter mass
companion with periastron $a (1 - e) \gta 2.55$ AU. However, even the
largest periastron value in this sample of systems is somewhat lower,
$\sim 2.5$ AU for HD38529, suggesting that long term stability may be
a problem (a host of papers have considered long term stability of
observed extrasolar planetary systems, including {\'E}rdi et al. 2004;
Menou \& Tabachnik 2003; Noble et al. 2002; Jones et al. 2001; and
many others). Indeed, simulations of extrasolar planetary systems with
added massless test particles (Barnes \& Raymond 2004) indicate that
small values of forced eccentricity (Figs.  \ref{fig:reshd38529} --
\ref{fig:reshd37124}) are necessary but not sufficient; these
simulations show that HD74156 and HD168443 eject almost all of the
test particles in a few Myr, while HD38529 and HD37124 allow for
continued particle survival. 

\subsubsection{Mean yearly flux versus eccentricity} 

An important (unresolved) question is the maximum allowed eccentricity
for a ``habitable'' planet. Such planets are generally assumed to have
nearly circular orbits. In planetary systems with detectable giant
planets, however, any possible terrestrial planets are likely to be
subject to eccentricity forcing and hence nonzero mean values of $e$.
Although the details of habitability must ultimately depend on the
characteristics of the planetary atmosphere, and such a discussion is
beyond the scope of this paper, the variations in the stellar flux are
straightforward to determine.  For nonzero eccentricities, the flux
received by the planet cannot be described by a single value, like the
well-known value of our solar constant 1.36 $\times 10^6$ erg
sec$^{-1}$ cm$^{-2}$, but rather a distribution of values. We can
determine the parameters of this distribution. Let $F_0$ be a fiducial
value of the flux defined by 
\be 
F_0 \equiv {L_\ast \over 4 \pi a^2} \, . 
\label{eq:fzero} 
\ee 
In other words, $F_0$ is the value of the (constant) flux received 
by a planet on a circular orbit of radius $a$. The extrema of the 
distribution are then given by $F_{max} = F_0 / (1 - e)^2$ and
$F_{min} = F_0 / (1 + e)^2$.  If we define a reduced flux 
$f = F/F_0 = (a/r)^2$, the distribution of reduced flux can  
be written in the form 
\be
{dP \over df} = {1 \over 2 \pi f^{3/2} } \bigl[ 
2 \sqrt{f} - 1 - f (1-e^2) \bigr]^{-1/2} \, . 
\label{eq:dpdf} 
\ee
The time-averaged mean value of the flux is then given by  
\be
\fbar = {F_0 \over (1 - e^2)^{1/2} } \, . 
\ee 
The requirement of a given mean flux (averaged over the planet's year)
thus demands that the quantity $a^2 \sqrt{1 - e^2}$ attain a particular
value, in contrast to the particular value of $a$ usually required for
circular orbits. Notice also that with increasing eccentricity, the
mean flux is larger for a given semi-major axis, or, equivalently, the
planet can live farther from its star. The variance (width) $\sigma_F$
of the distribution can be written in the form
\be
\sigma_F = \fbar 
\Bigl[ {1 + e^2/2 \over (1 - e^2)^{3/2} } - 1 \Bigr]^{1/2} \, , 
\label{eq:sigmaf} 
\ee
so that the distribution of flux experienced by the planet grows wider
with increasing eccentricity (as expected). 

Highly eccentric orbits will produce large seasonal variations on the
planet (in addition to those produced by possible tilting of the spin
axis) and it remains to be determined how large such variations can be
and still allow for a habitable world. Sufficiently large
eccentricities lead to nonlinear seasonal flux variations. One useful
benchmark occurs for the case $\sigma_F = \fbar$, where the flux
distribution is as wide as its mean, which in turn occurs for
eccentricity $e_0 \approx 0.554$.  With this eccentricity, the yearly
flux varies by a factor of $\sim$12 from maximum to minimum. As
another benchmark, even a modest eccentricity of $e \approx 0.172$
will enforce a factor of two variation in flux over the planetary
orbit. Although the flux variations are straightforward to calculate,
the degree to which planets can maintain stable climates in the face
of such variations remains to be determined.  In any case, these
considerations have important implications for astrobiology (e.g., 
Lunine 2005).  

Keep in mind that the distributions of flux considered here are the
possible values of flux experienced by the planet over the course of
its ``year'' for a given eccentricity. Over longer time scales, those
determined by secular interactions, the eccentricity itself will
sample a distribution of values as discussed in the previous
subsection.

\subsection{General Relativity in Extrasolar Planetary Systems} 

The combination of secular interactions and general relativity
provides a number of interesting results. For some extrasolar
planetary systems, one can use observed system parameters to provide
new tests of general relativity; in other systems, one can use general
relativity to place new physical constraints on the system parameters 
(see also Adams \& Laughlin 2006b). 

In order for the effects of general relativity to be significant, the
final term in equation (\ref{eq:diag}) must complete with the others.
In practice, only the inner planet will have significant corrections.
To find an analytic criterion for the importance of relativistic
effects, we assume that the system is sufficiently hierarchical so
that $b^{(1)}_{3/2}(\alpha_{j0}) = 3 \alpha_{j0} + {\cal O}
(\alpha_{j0}^2) \approx 3 \alpha_{j0}$ and thus obtain the requirement
\be 
\sum_j {m_j \over M_\ast} \alpha_{j0}^3 \sim {4 G M_\ast \over c^2 a_0} 
\, .
\ee 
Both sides of this equation are small dimensionless quantities;
however, when their ratio is of order unity, then relativistic effects
can compete with secular interactions.  Since in most cases, only one
of the (non-inner) planets will contribute to the sum on the left hand
side of the equation, this constraint is equivalent to the requirement
that one of the dimensionless fields $\Pi_j \gta 1$, where 
\be 
\Pi_j \equiv {4 G M_\ast^2 a_j^3 \over c^2 m_j a_0^4 } \, \approx 
6.3 \Bigl( {m_j \over m_J} \Bigr)^{-1} 
\Bigl( {M_\ast \over M_\odot} \Bigr)^{2} 
\Bigl( {a_j \over 1 \, {\rm AU} } \Bigr)^{3} 
\Bigl( {a_0 \over 0.05 \, {\rm AU}} \Bigr)^{-4} \, . 
\label{eq:pidef} 
\ee  
The second (approximate) equality indicates that relativistic effects
compete with non-relativistic secular interactions for a Jupiter mass
planet in a 1 AU orbit perturbing a hot Jupiter (in a 4 day orbit).
The relative size of the relativistic effect grows with increasing
semi-major axis $a_j$ of the perturber, but the magnitude of the
secular contribution decreases.

This problem also contains a characteristic time scale $\taugr$ defined by 
\be
\taugr \equiv {c^2 a_0^{3/2} \over 3 (G M_\ast)^{3/2}} \approx 3011 \, 
{\rm yr} \, \Bigl( {a_0 \over 0.05 \, {\rm AU} } \Bigr)^{5/2} \, 
\Bigl( {M_\ast \over 1.0 M_\odot } \Bigr)^{-3/2} \, . 
\ee 
This scale represents the time required for the periastron of a hot
Jupiter (in about a 4-day orbit) to precess one radian forward in its
orbit. Notice that this time scale is comparable to the secular
interaction time scales (determined by the Laplace-Lagrange
eigenvalues) shown in Table 1 for many of the observed systems. This
concordance of time scales, $\taugr \sim \tau_{\rm sec}$, is purely
accidental (in that a different value of the speed of light $c$ would
change the result, and planetary system formation probably does not
depend on relativistic effects) but allows for GR to play a
significant role in the dynamics of some planetary systems.

To illustrate the action of relativity in extrasolar planetary
systems, we have calculated the secular interactions for two systems
with and without including the general relativistic terms. We use the
systems Upsilon Andromedae and HD160691 because they contain hot
Jupiters as well as additional planets in orbits with $a \sim 1$
AU. Figure \ref{fig:genrel} shows the mean eccentricity $\ebar$ of the
innermost planet, as driven by secular interactions and averaged over
many cycles, plotted as a function of $\sin i$ for the two cases.

For the Upsilon Andromedae system (top panel of Fig. \ref{fig:genrel}), 
the inclusion of relativity acts to damp the excitation of
eccentricity by secular interactions.  For large values of $\sin i$,
the innermost planet would be driven to eccentricity values $\ebar
\approx 0.4$ without relativity, but only $\ebar \approx 0.016$ when
the relativistic corrections are included.  The observed eccentricity
for the inner planet in this system is $e_{obs} \sim 0.011$, much
closer to the relativistic mean value. As a result, one can think of
Upsilon Andromedae as providing another test of general relativity.
Since secular theory uses the current (observed) eccentricity values
of the planets as part of the boundary conditions, the system always
has a chance of displaying the observed (low) eccentricity of the
inner planet, so the implications for general relativity must be
stated in terms of probabilities: If we assume that the observed
eccentricity of the inner planet has a measurement error that is
gaussian distributed with $\sigma_{obs} = 0.015$, then the probability
of observing the system in its current state would be only 0.0235 if
the general relativistic corrections are not included. An analogous
calculation implies a probability of 0.78 of finding the inner planet
with its observed eccentricity if general relativity is correct. The
validity of GR is thus strongly favored.

For the HD160691 system (bottom panel of Fig. \ref{fig:genrel}), the
inclusion of relativistic terms acts in the opposite direction, i.e.,
it leads to greater predicted values of $\ebar$. For large values of
$\sin i \sim 1$, the predicted mean driven eccentricities are small
enough to be consistent with the observed low value $e_{obs} \sim
0$. As the value of $\sin i$ decreases, however, the level of
eccentricity forcing increases and reaches a resonance near $\sin i
\approx 0.5$. The observed low value of eccentricity, in conjunction
with these considerations, thus constrain the viewing angle of the
HD160691 system to be nearly edge-on; if we require $\ebar \lta 0.05$,
then the viewing angle is confined to the range $\sin i \gta 0.93$ ($i
\gta 70^\circ$). Since inclination angles are notoriously difficult
to measure in these systems, this constraint on $i$ is quite valuable.

This constraint on the viewing angle also has implications for the
possibility of observing the inner planet in transit.  For a hot
Jupiter in a zero eccentricity orbit, the {\it a priori} probability
${\cal P}_0$ that the planet can be seen in transit is given by
\be
{\cal P}_0 = {\int_0^{(R_\ast + r_P)/a} d (\cos i) \over 
\int_0^1 d (\cos i) } = {R_\ast + R_P \over a} \,  . 
\ee 
For the HD160691 system, the inclination angle is confined to the
narrower range $\sin i \gta 0.93$ ($\cos i \lta 0.37$) and the
probability that the inner planet can be seen in transit is larger
than the {\it a priori} value by a factor of 2.7. With this constraint
on the inclination angle, the probability of finding observable
transits in HD160691 is ${\cal P} \sim 0.14$, making it one of the
higher probability systems\footnote{www.transitsearch.org}.
 
Most previous discussions of relativistic effects in planetary systems
emphasize its stabilizing influence. However, general relativity can
lead to either larger eccentricity forcing (HD160691) and hence less
stability, or smaller eccentricity forcing (Ups And) and hence greater
stability (Adams \& Laughlin 2006b).  It is useful to have an analytic
criterion for when the two types of behavior occur, and what system
parameters are required for GR to provide more/less stability. 

For a two planet system, we can isolate the effects of relativity by
considering the simplest case in which the inner planet has a nearly
circular orbit, or at least cycles through the $e_1$ = 0 state. This
case is relatively common in that close planets tend to have nearly
circular orbits and close planets display relativistic effects.  The
solution of equation (\ref{eq:solution}) can be used to find the
square of the eccentricity of the inner planet. After time averaging,
the resulting amplitude can be written in terms of the eigenvectors so
that $\langle e_1^2 \rangle = \eta^2 = \Lambda_{11}^2 +
\Lambda_{12}^2$. By finding the eigenvectors in terms of the original
matrix elements $A_{ij}$ and applying the boundary conditions, we can
write the eccentricity amplitude of the inner planet (as forced by the
outer planet) in the form 
\be
\eta^2 = {A_{12}^2 e_{02}^2 \over (\lambda_1 - \lambda_2)^2 } \, ,   
\label{eq:etadef} 
\ee
where $e_{02}$ is the initial eccentricity of the second (outer)
planet (see the following subsection).  For a two planet system, 
the difference in eigenvalues is given by the expression  
\be 
\lambda_1 - \lambda_2 = 
\bigl[ (A_{11} - A_{22})^2 + 4 A_{12} A_{21} \bigr]^{1/2} \, ,  
\ee 
which is positive definite.  Notice that the eigenvalues cannot be
degenerate for a two planet system, and hence secular interactions do
not lead to resonance.  Notice also that only the diagonal matrix
elements contain relativistic corrections, and only the inner planet
is generally close enough to the star for such corrections to make a
difference. The first matrix element can thus be written 
\be 
A_{11} = A_0 + 3 n_1 \mu \, , 
\ee 
where $A_0$ is the matrix element in the absence of relativity and
$\mu = G M_\ast/(c^2 a_1)$ is the dimensionless relativistic
correction factor. With this construction, the question of whether
relativity acts to enforce larger or smaller amplitudes of
eccentricity oscillation is determined by the sign of the derivative 
$d \eta / d\mu$.  In order to outline its dependence on the system 
parameters, we make the same approximations as before ($m_1, m_2 \ll
M_\ast$; $a_1 \ll a_2$ so that $b_{3/2}^{(1)} \approx 3 a_1/a_2$;
etc.) and write the derivative in the form
\be
\sgn \Big( {d \eta \over d \mu} \Bigr) = - 
\sgn \Big[ 1 + \Pi - {m_1 \over m_2} 
\big( {a_1 \over a_2} \bigr)^{1/2} \Big] \, , 
\ee 
where $\Pi$ is the dimensionless field defined in equation
(\ref{eq:pidef}). The third term must dominate in order for the sign
to be positive, and hence for relativity to lead to greater
eccentricity amplitudes.  Since $a_1 < a_2$ by definition, and since
the inner planets tend to be less massive ($m_1 < m_2$), relativity
will usually lead to greater stability. For sufficiently massive inner
planets, however, relativity can amplify eccentricity forcing. One 
convenient way to write the constraint (requirement) for eccentricity 
amplification is in the form 
\be 
m_2 \, < \, m_1 \big( {a_1 \over a_2} \bigr)^{1/2} 
- 4 \mu M_\ast \big( {a_1 \over a_2} \bigr)^{-3} \, . 
\label{eq:mrellimit} 
\ee 
Keep in mind that we are working in the limit $a_1 \ll a_2$, which is
also required for long term dynamical stability. As a result, this
constraint cannot, in practice, be satisfied by making the ratio
$a_1/a_2$ too large; instead, the mass $m_2$ must be small compared
with $m_1$. The two planet systems considered in this paper (Table 2)
show an interesting trend: None of the systems satisfy the inequality
of equation (\ref{eq:mrellimit}), so that general relativity leads to
greater stability for all of these systems.

Now consider a three planet system in which the inner planet has
relatively little effect on the outer two planets, due to its smaller
mass and/or inner position (far from the others). This description
applies to both HD160691 and UpsAnd. In this hierarchical limit, the
$3 \times 3$ matrix $A_{ij}$ can be approximated by taking $A_{12} =
A_{13} = A_{21} = A_{31} = 0$. The eigenvalues of this reduced
matrix then take the form 
\be
\lambda_1 = A_{11} \, , 
\lambda_{2,3} = {1 \over 2} \bigl\{ (A_{22} + A_{33} \pm 
\bigl[ (A_{33} - A_{22})^2 + 4 A_{23} A_{32} \bigr]^{1/2} 
\bigr\} \, . 
\label{eq:gr3} 
\ee 
With this level of complexity, the system can have degenerate
eigenvalues, e.g., when $\lambda_1 = A_{11}$ is equal to either
$\lambda_2$ or $\lambda_3$.  Without loss of generality, suppose that
$\lambda_1$ and $\lambda_2$ are nearly degenerate. The effect of
general relativity is to add a small positive contribution to $A_{11}$
and hence $\lambda_1$ (we again take $A_{11} = A_0 + 3 n_1 \mu$),
where the physical implication of the added term is the general
relativistic precession of the orbit of the inner planet. If
$\lambda_1 \lta \lambda_2$, then GR brings the eigenvalues closer
together and thus acts to increase the level of eccentricity
forcing. If $\lambda_1 \gta \lambda_2$, then GR makes the eigenvalues
more unequal and acts to decrease eccentricity forcing. The full cubic
equation for $\det [A - \lambda I]$ is more complicated, but allows
for similar behavior. 

The basic result of this section is that general relativity can have a
significant effect on secular perturbations for planetary systems with
favorable characteristics (see also Holman et al. 1997; Ford et
al. 2000ab). The requirement that GR effects are important is
quantified by the dimensionless field defined in equation
(\ref{eq:pidef}). Furthermore, general relativity can either enhance
or attenuate the effects of secular perturbations, where the sign of
the relativistic effects is described by equations (\ref{eq:etadef} --
\ref{eq:gr3}).

\subsection{Constraints on Additional Planets in Systems with Hot Jupiters} 

Most hot Jupiters are observed to have small or zero eccentricity
values and relatively low mass. For those systems that have observed
hot Jupiters but no additional planets detected (thus far), the theory
of secular interactions places a clean constraint on any possible
companions. The presence of any additional planet (or larger body)
will tend to excite eccentricity in the hot Jupiter. For the case of
interest, we can set $e_1(t=0)$ = 0 (the hot Jupiter is observed in a
circular orbit) and $\varpi_2 (t=0)$ = 0 (we choose the orientation of
the coordinate system so that the possible companion planet has zero
argument of periastron). With these simplifications, the enforced
eccentricity of the inner planet (the hot Jupiter) samples a
distribution characterized by the following parameters: 
\be 
e_{min} = 0 \, , \qquad e_{max} = 2 \eta \, , \qquad 
\ebar_1 = {4 \over \pi} \eta \, , \qquad 
\langle e_1^2 \rangle = 2 \eta^2 \, , \qquad {\rm and} 
\qquad \sigma_1^2 = \bigl[2 - 16/\pi^2\bigr] \eta^2 \, , 
\ee
where the amplitude $\eta$ is defined in terms of the eccentricity
$e_{2(obs)}$ of the second planet and the matrix elements $A_{ij}$,
i.e., 
\be 
\eta^2 \equiv {A_{12}^2 \over (A_{11} - A_{22})^2 + 4 A_{12} A_{21}} 
\, \, e_{2(obs)}^2 \, \, .  
\ee 
With this formulation, we can find the orbital elements of a second
planet that would drive the eccentricity of the hot Jupiter to a
specified mean value $\ebar_1$.  Figure \ref{fig:hotjup} shows the
resulting $a-e$ plane for the orbital elements of the (hypothetical)
second planet. For example, the lowest curve shows the required
eccentricity $e_2$ of the second planet as a function of its
semi-major axis $a_2$ such that the inner planet is excited to have a
mean eccentricity $\ebar_1$ = 0.01. The shaded region below this curve
depicts the portion of parameter space for which the second planet has
a negligible effect on the hot Jupiter.  The corresponding curves for
$\ebar_1$ = 0.02, 0.05, 0.10, 0.20, and 0.50 are also shown. The
observed hot Jupiters do not have companions in the unshaded region of
the plane.  However, such planets could exist -- the square symbols in
Figure \ref{fig:hotjup} depict the locations of known planets in the
$a-e$ plane. Notice that the relevant portion of the $a-e$ plane is
relatively small. The curves shown in Figure \ref{fig:hotjup} assume
that the hot Jupiter has a period of 4 days and the minimum period of
the second planet is 12 days, just outside the 3:1 mean motion
resonance (i.e., wide enough separations so that we expect secular
theory to apply).  The eccentricity $e_2$ required to drive a given
$\ebar_1$ grows rapidly with increasing semi-major axis (period) and
exceeds unity for relatively modest values of $a_2$ as shown in the
figure. Of course, the eccentricity of the inner planet cycles through
a distribution of values. For this case, the cumulative probability
${\cal P} (e)$ that the inner planet is observed with an eccentricity
$e_{obs} < e$, for a system with mean (forced) eccentricity $\ebar_1$,
is given by the expression 
\be 
{\cal P} (e) = {2 \over \pi} \sin^{-1}
\Bigl( {2 e \over \pi \ebar_1} \Bigr) \, .  
\ee
This form can be derived by evaluating the probability distribution of
equation (\ref{eq:dpde}) for the case under consideration (where
$\Lambda_{11} = \Lambda_{12} = \eta$) and performing the integral to
find the cumulative probability.

\section{Conclusion} 

\begin{figure} 
\figurenum{1} 
\centerline{ \epsscale{0.9} \plotone{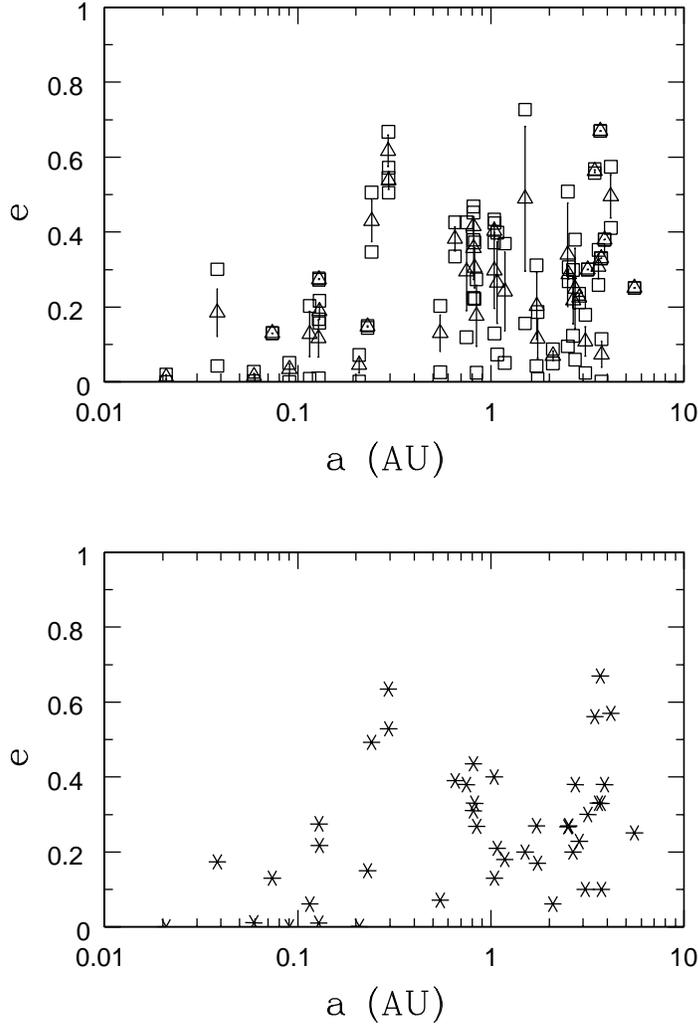} } 
\caption{ The effects of secular interactions on the $a-e$ plane
for observed multiple planet systems. The bottom panel shows the $a-e$
plane using the observed eccentricities for all of the planets found
in multiple systems (to date). The top panel shows the eccentricities
predicted by Laplace-Lagrange secular theory for the same planets. The
mean values of the eccentricity, averaged over many secular cycles,
are shown as open triangles. The errorbars depict the range of
eccentricities over the same cycles, as measured by the variance of
the distribution, and the open squares depict the minimum and maximum
values. }
\label{fig:aeplane} 
\end{figure}

\begin{figure} 
\figurenum{2} 
\centerline{ \epsscale{0.80} \plotone{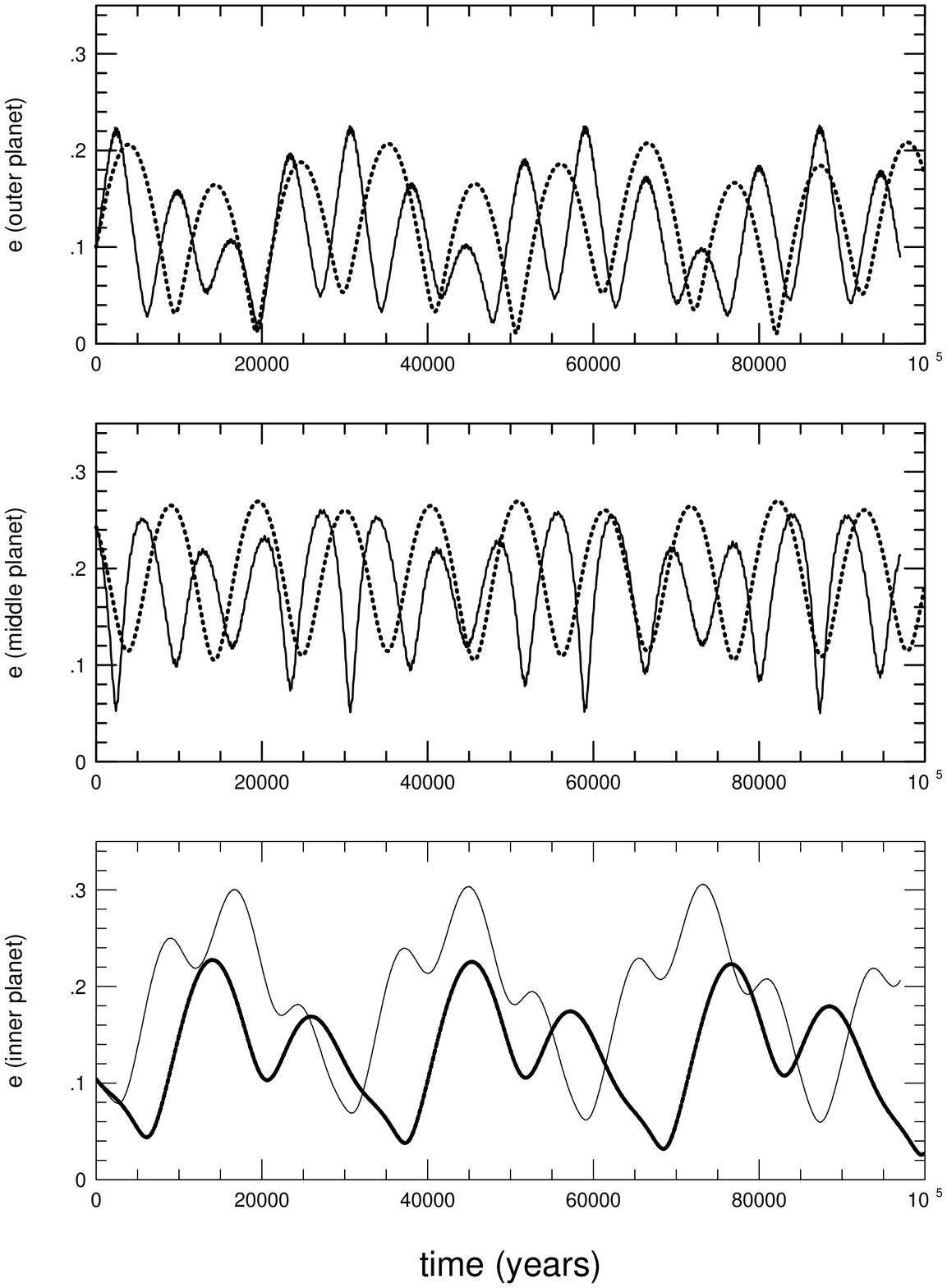} } 
\caption{ Comparison of secular theory with direct numerical
integration for extrasolar planetary system HD37124. The eccentricity
variations produced by direct numerical integration are shown by the
solid curves; the corresponding eccentricity variations predicted by
secular theory are shown by the dotted curves. The two approaches are
in good agreement. }
\label{fig:numhd37124} 
\end{figure}

\begin{figure} 
\figurenum{3} 
\centerline{ \epsscale{0.80} \plotone{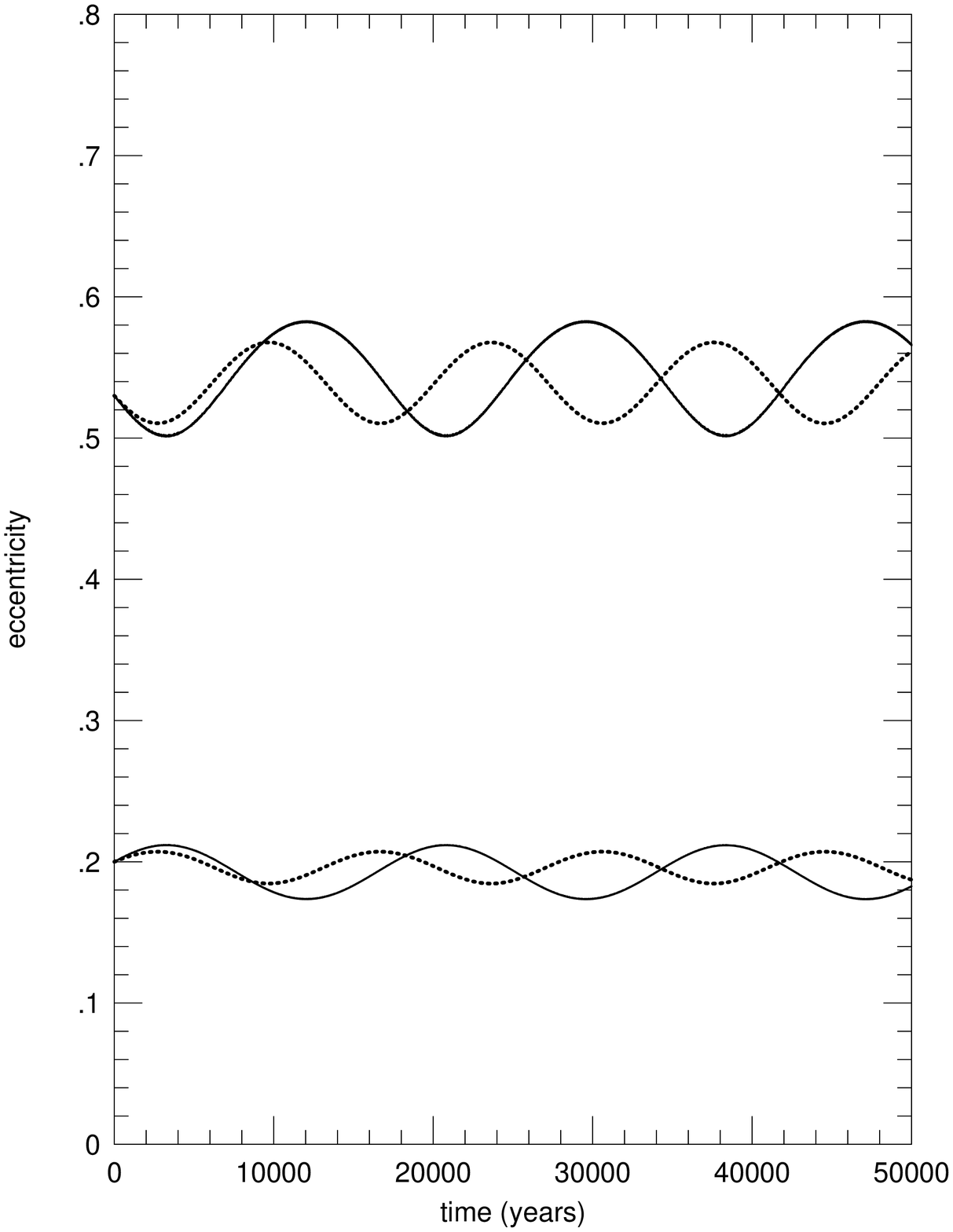} } 
\caption{ Comparison of secular theory with direct numerical
integration for extrasolar planetary system HD168443. The eccentricity
variations produced by direct numerical integration are shown by the
solid curves; the corresponding eccentricity variations predicted by
secular theory are shown by the dotted curves. The two approaches are 
in good agreement. } 
\label{fig:numhd168443} 
\end{figure}

\begin{figure} 
\figurenum{4} 
\centerline{ \epsscale{0.80} \plotone{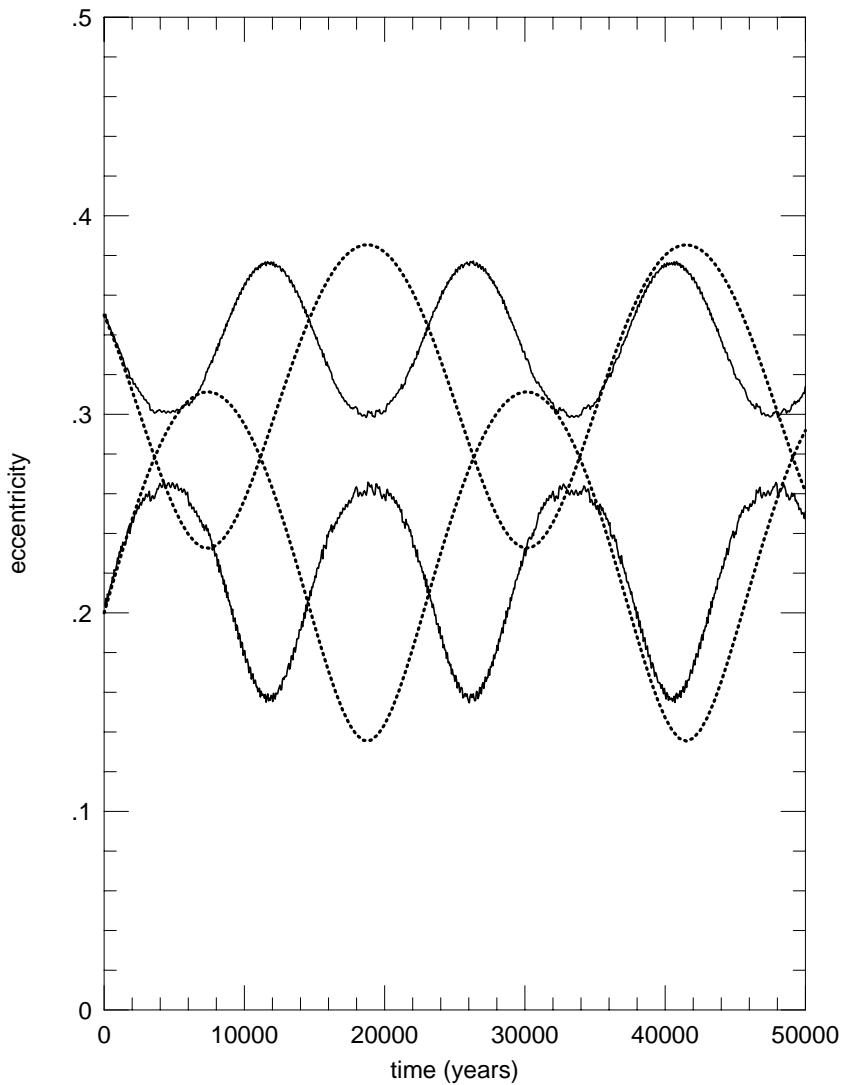} } 
\vskip-0.25truein 
\caption{ Comparison of secular theory with direct numerical
integration for extrasolar planetary system HD12661. The dotted curves 
show the result of the secular theory discussed in the text. The two
sets of solid curves show the result of direct numerical integration 
using two different sets of starting parameters (both within the
observational uncertainties). The results show considerable variation
between the two sets of starting parameters and between the numerical
and secular integrations. Nonetheless, the overall envelope of
eccentricity variation -- and hence the distribution of eccentricities
sampled by the planets -- is similar for the cases shown here. }  
\label{fig:numhd12661}  
\end{figure}

This paper has applied the leading order theory of secular
interactions to observed extrasolar planetary systems and used the
results to constrain their properties. Our results can be summarized
as follows:

[1] Through dynamical interactions, described here using secular
theory, the orbital eccentricities in multiple planet systems vary
over secular time scales.  The eccentricities measured by ongoing
planet searches represent the current eccentricity value, which is
drawn from a wider distribution of values sampled by the planet. In
other words, the eccentricities in multiple planet systems should not
be considered as particular values, but rather as distributions of
values.  The widths of these eccentricity distributions can be
substantial (Tables 2 and 3) and we have verified that secular theory
predicts distribution widths that are in good agreement with direct
numerical integration (Figures \ref{fig:numhd37124} --
\ref{fig:numhd12661}). This effect implies that interpretations of the
$a-e$ diagram must be regarded with caution for multiple planet
systems (see Figure \ref{fig:aeplane}). For the simplest case of two
planet systems, the resulting distribution of eccentricity can be
found analytically (eqs. [\ref{eq:dpde} -- \ref{eq:sigmae}]).  The
time scale for secular eccentricity variations is typically thousands
of years (Table 1), much longer than observational survey time scales
(tens of years) and much shorter than the system lifetimes (few Gyr).

\begin{figure}
\figurenum{5}  
\centerline{ \epsscale{0.9} \plotone{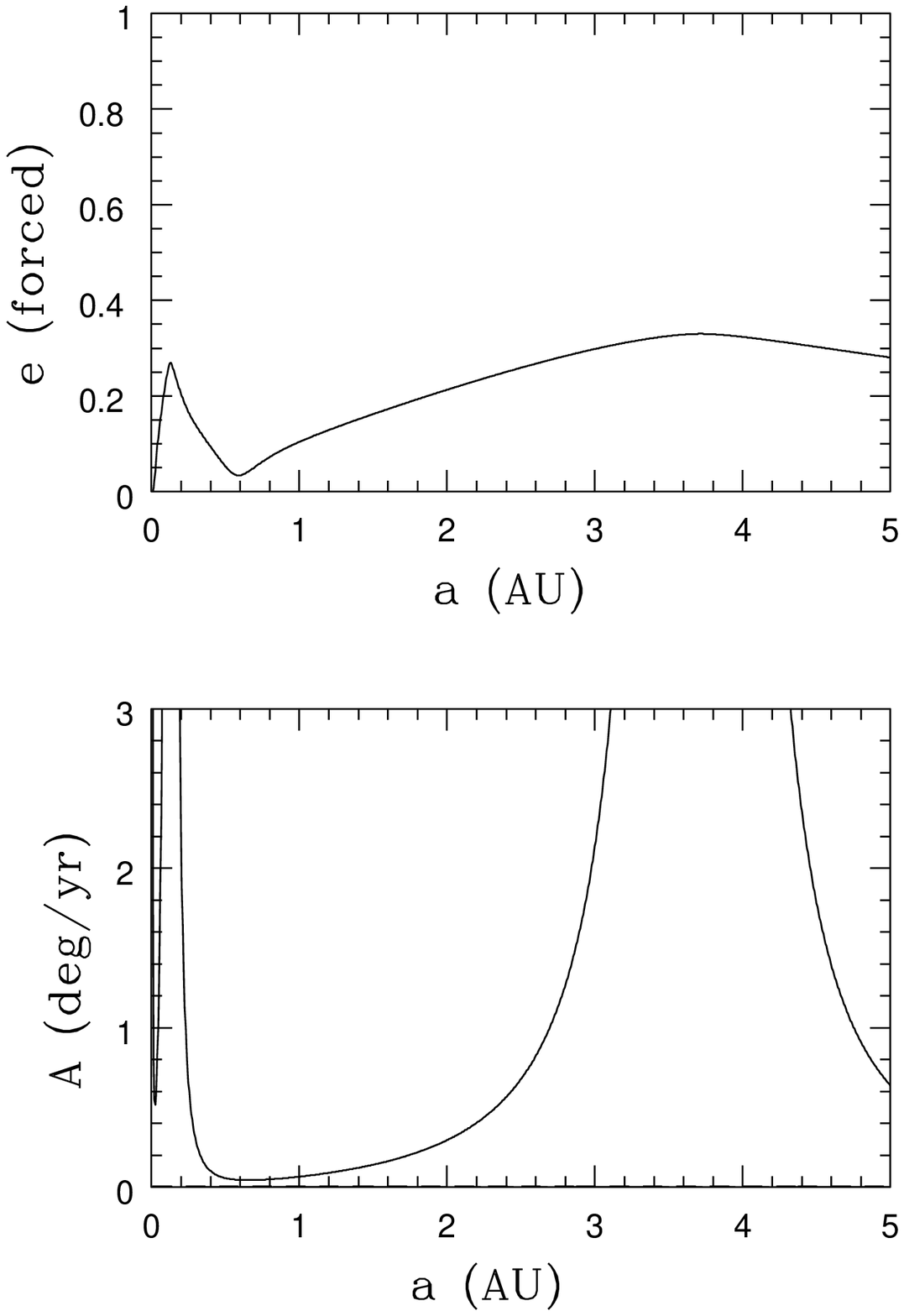}} 
\caption{ Effects of secular interactions on a hypothetical test
particle in the observed planetary system HD38529. The bottom panel
shows the oscillation frequency $A$ of the test particle as a function
of its semi-major axis $a$. The values of the eigenfrequencies 
$\lambda_1, \lambda_2$ for variations in eccentricity/periastron fall 
well below the solid curve. The top panel shows the forced 
eccentricity as a function of $a$ for time $t$ = 0. }
\label{fig:reshd38529} 
\end{figure}

\begin{figure} 
\figurenum{6}  
\centerline{ \epsscale{0.9} \plotone{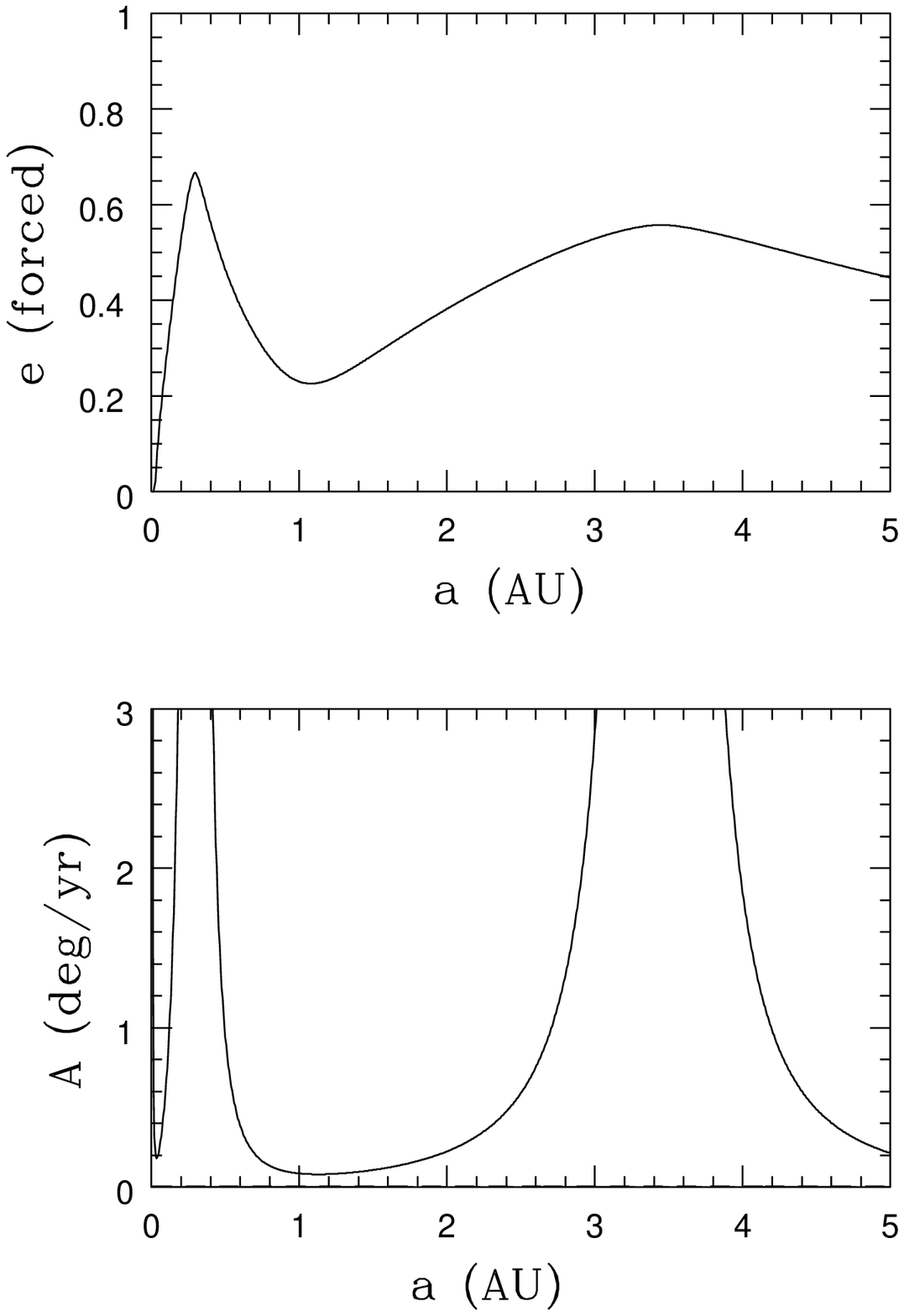}} 
\caption{ Effects of secular interactions on a hypothetical test
particle in the observed planetary system HD74156. The bottom panel
shows the oscillation frequency $A$ of the test particle as a function
of its semi-major axis $a$. The values of the eigenfrequencies 
$\lambda_1, \lambda_2$ for variations in eccentricity/periastron fall 
well below the solid curve. The top panel shows the forced
eccentricity as a function of $a$ for time $t$ = 0.}
\label{fig:reshd74156} 
\end{figure}

\begin{figure}
\figurenum{7}   
\centerline{ \epsscale{0.9} \plotone{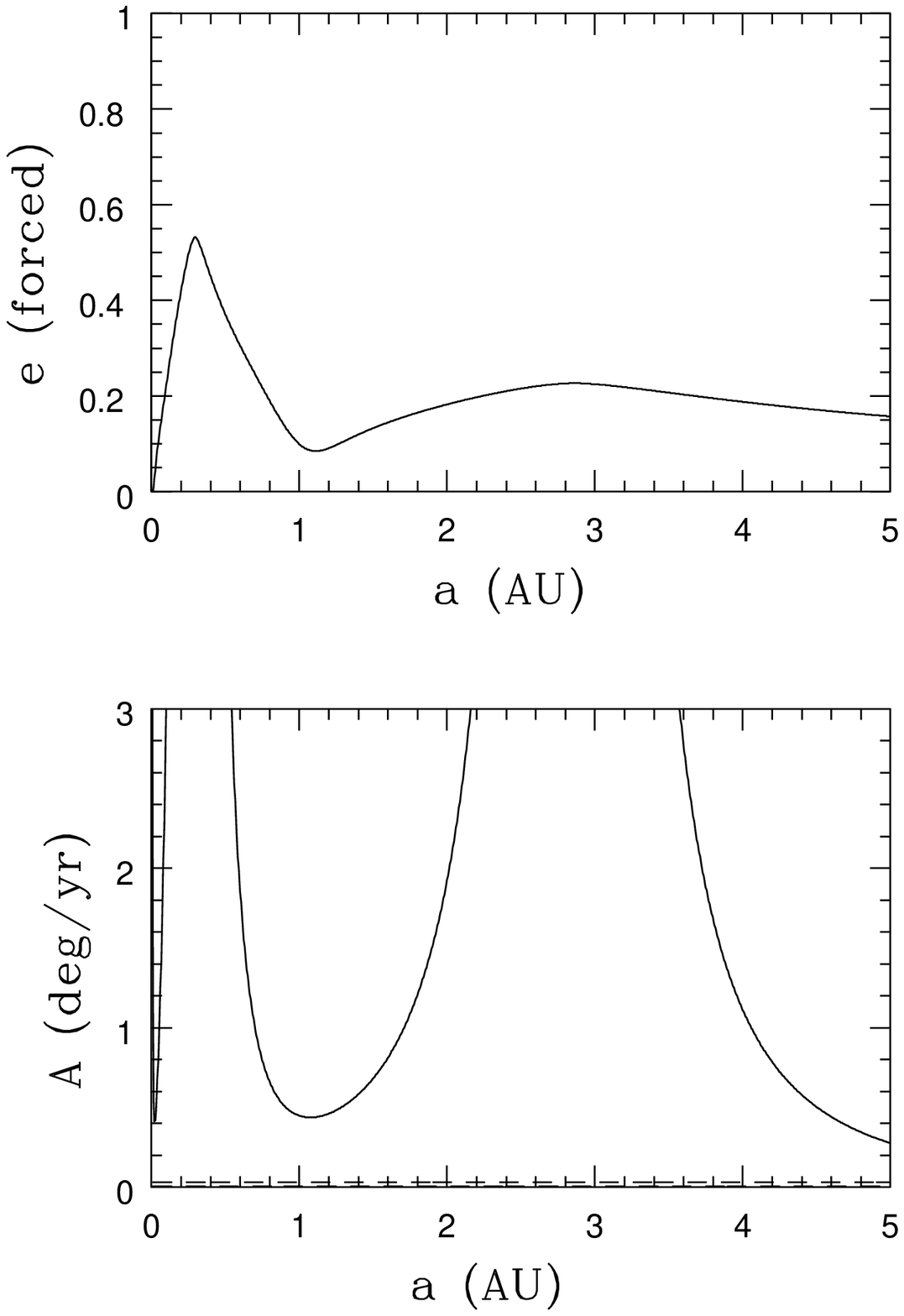}} 
\caption{ Effects of secular interactions on a hypothetical test
particle in the observed planetary system HD168443. The bottom panel
shows the oscillation frequency $A$ of the test particle as a function
of its semi-major axis $a$. The values of the eigenfrequencies 
$\lambda_1, \lambda_2$ for variations in eccentricity/periastron 
(dashed lines) fall well below the solid curve. The top panel shows 
the forced eccentricity as a function of $a$ for time $t$ = 0.} 
\label{fig:reshd168443} 
\end{figure}

\begin{figure}
\figurenum{8}   
\centerline{ \epsscale{0.9} \plotone{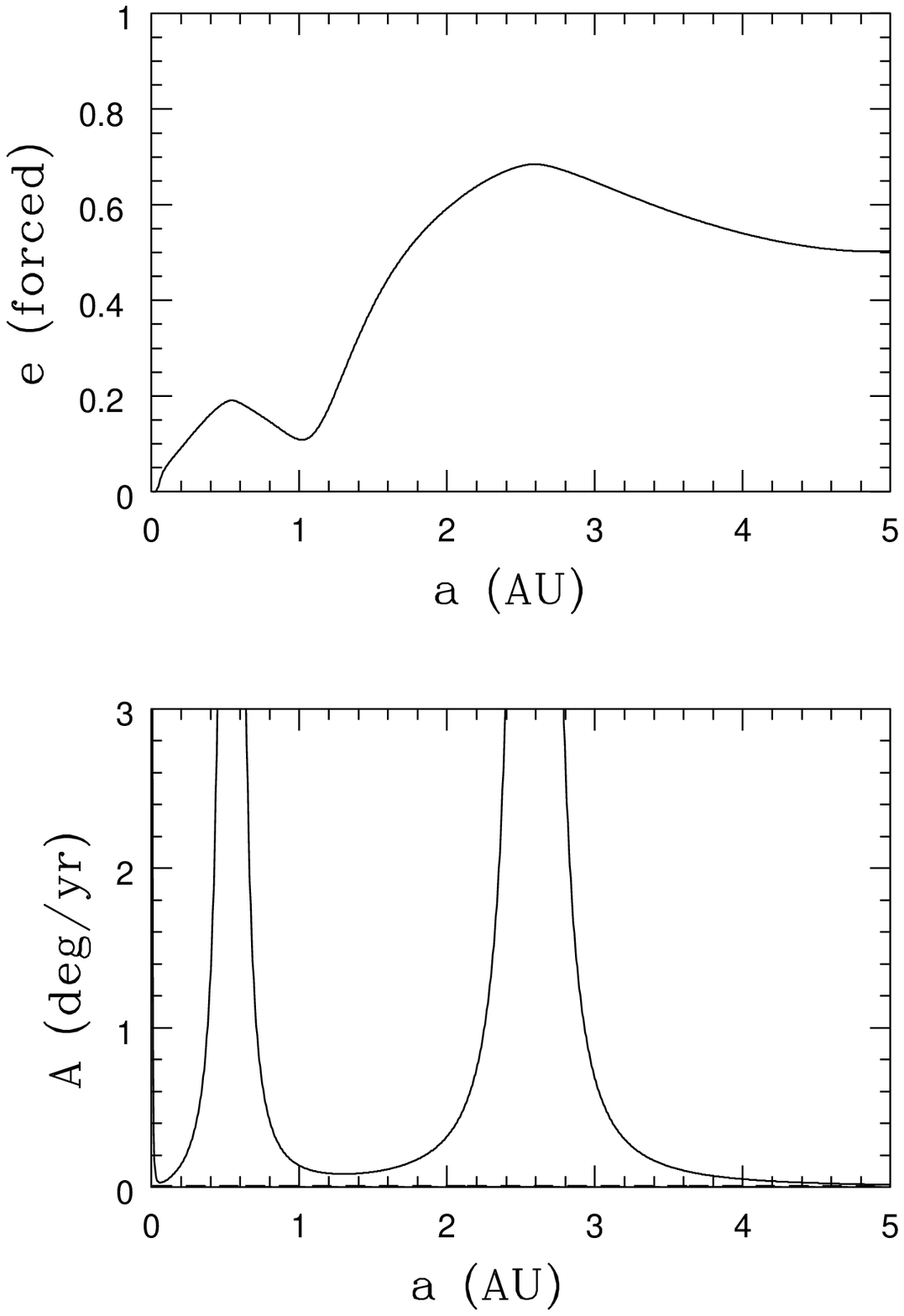}} 
\caption{ Effects of secular interactions on a hypothetical test
particle in the observed planetary system HD37124. The bottom panel
shows the oscillation frequency $A$ of the test particle as a function
of its semi-major axis $a$. The values of the eigenfrequencies 
$\lambda_1, \lambda_2$ for variations in eccentricity/periastron fall 
well below the solid curve. The top panel shows the forced
eccentricity as a function of $a$ for time $t$ = 0.}
\label{fig:reshd37124} 
\end{figure}

\begin{figure} 
\figurenum{9}  
\centerline{ \epsscale{0.9} \plotone{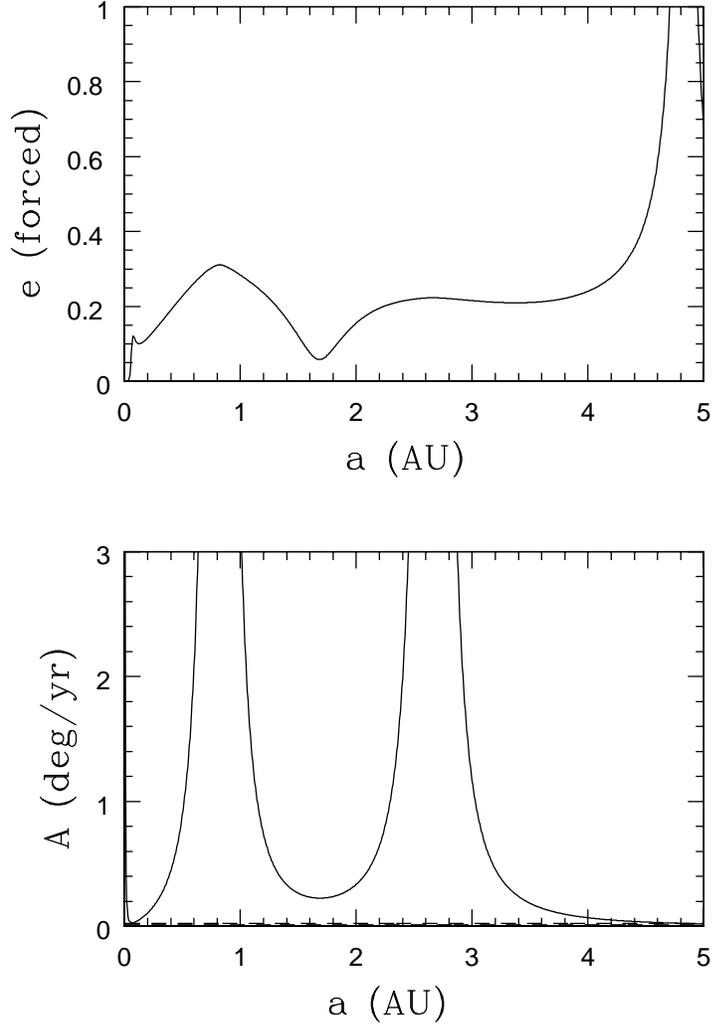}} 
\caption{ Effects of secular interactions on a hypothetical test
particle in the observed planetary system HD12661. The bottom panel
shows the oscillation frequency $A$ of the test particle as a function
of its semi-major axis $a$. The dashed horizontal lines denote the
values of the eigenfrequencies $\lambda_1, \lambda_2$ for variations
in eccentricity/periastron. The top panel shows the forced
eccentricity as a function of $a$ for time $t$ = 0.}
\label{fig:reshd12661} 
\end{figure}

[2] For a sub-sample of observed multiple planet systems, we have
calculated the frequencies of free oscillation and amplitudes of
forced eccentricity oscillation (Figures \ref{fig:reshd38529} --
\ref{fig:reshd12661}) for (hypothetical) test bodies. These results
constrain the possibility of multiple planet systems containing
additional small (terrestrial) planets. This calculation shows that
many of the systems allow the existence of additional planets, but the
forced eccentricity amplitudes would be large compared to that of
Earth, i.e., $e$(forced) $\approx$ 0.10 -- 0.15. In addition,
numerical integrations (Barnes \& Raymond 2004) indicate that only a
fraction of these systems allow test bodies to remain stable over long
time scales, i.e., the condition of small forced eccentricity
amplitude is necessary but not sufficient.

[3] This paper generalizes the concept of a habitable zone to include
the possibility of larger eccentricity values (\S 3.2.2). Specifically,
we have calculated the distribution of flux experienced by a planet in
an eccentric orbit over the course of its year (eq. [\ref{eq:dpdf}]),
as well as the mean flux and the width of the distribution (eqs.
[\ref{eq:fzero} -- \ref{eq:sigmaf}]). Notice that all of these results
are calculated analytically. 

\begin{figure}
\figurenum{10}  
\centerline{ \epsscale{0.9} \plotone{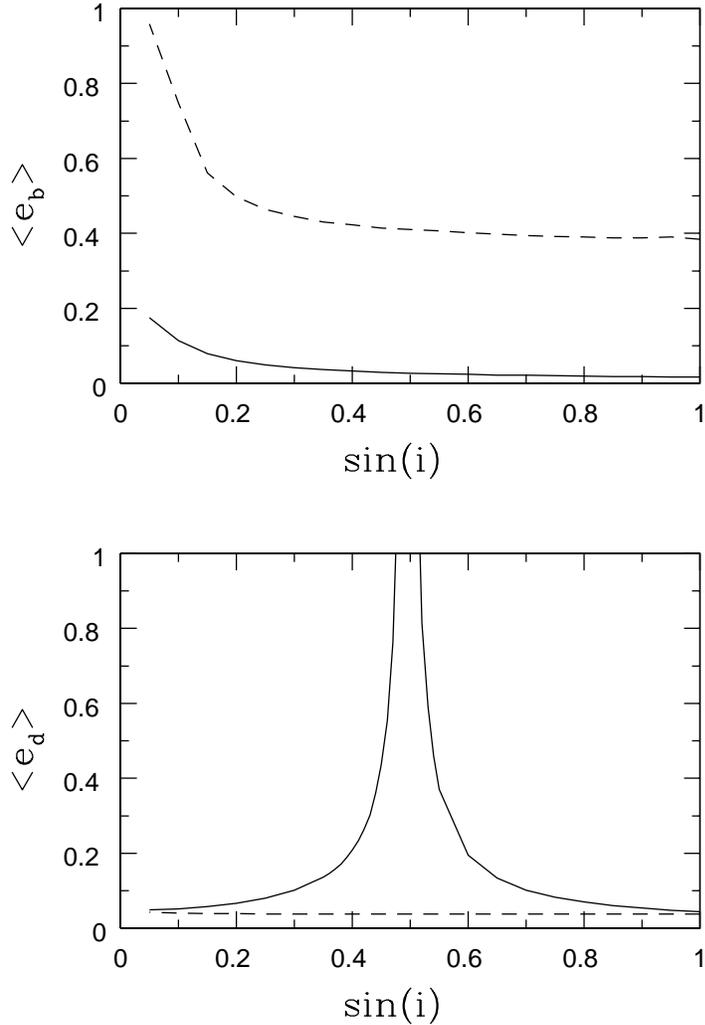} } 
\vskip-0.25truein 
\caption{ The effects of general relativistic corrections on two
extrasolar planetary systems. The mean eccentricity $\ebar$ of the
innermost planet, as driven by secular interactions, is plotted as a
function of $\sin i$ for the Upsilon Andromedae system (top panel) and
the HD160691 system (bottom panel). Both systems have three planets
detected to date. The predictions of secular theory are shown with
relativistic corrections as the solid curves, and without relativistic
corrections as the dashed curves. Notice that the relativistic terms
act in opposite ways in the two systems: Inclusion of relativity acts
to damp eccentricity excitation by the secular interactions in the Ups
And system. In the HD160691, however, they allow the system to
approach a resonance for $\sin i \approx 0.5$. }
\label{fig:genrel} 
\end{figure}

[4] For multiple planet systems with favorable architectures, the
effects of general relativity can be significant (Fig.
\ref{fig:genrel}). When planetary systems have two relatively massive
outer planets and a third inner planet of smaller mass near secular
resonance, the effects of general relativity are large enough to move
the system in or out of a resonant condition.  Since the resonance
condition depends on planetary masses, which in turn depend on $\sin
i$, and since small inner planets cannot survive in exact resonance,
this effect can be used to constrain the allowed range of $\sin i$ in
some observed extrasolar planetary systems. For the HD160691 system,
e.g., this constraint implies that $\sin i \gta 0.93$.  This effect
thus provides another means of probing the properties of these
planetary systems. For other systems (e.g., Upsilon Andromedae),
general relativity can make the inner planet precess forward in its
orbit fast enough to compromise eccentricity pumping from a second
planet, i.e., the mean eccentricity values are much smaller than they
would be in flat space. This latter effect can thus be used as a test
of general relativity. For Ups And, the system has a 78\% chance of
being observed with its measured parameters if general relativity is
correct, but only a $\sim 2\%$ chance if the general relativistic
corrections were zero. In the future, additional extrasolar planetary
systems and/or higher precision observations can provide more
stringent tests of GR (see also Adams \& Laughlin 2006b).

\begin{figure} 
\figurenum{11} 
\centerline{ \epsscale{0.8} \plotone{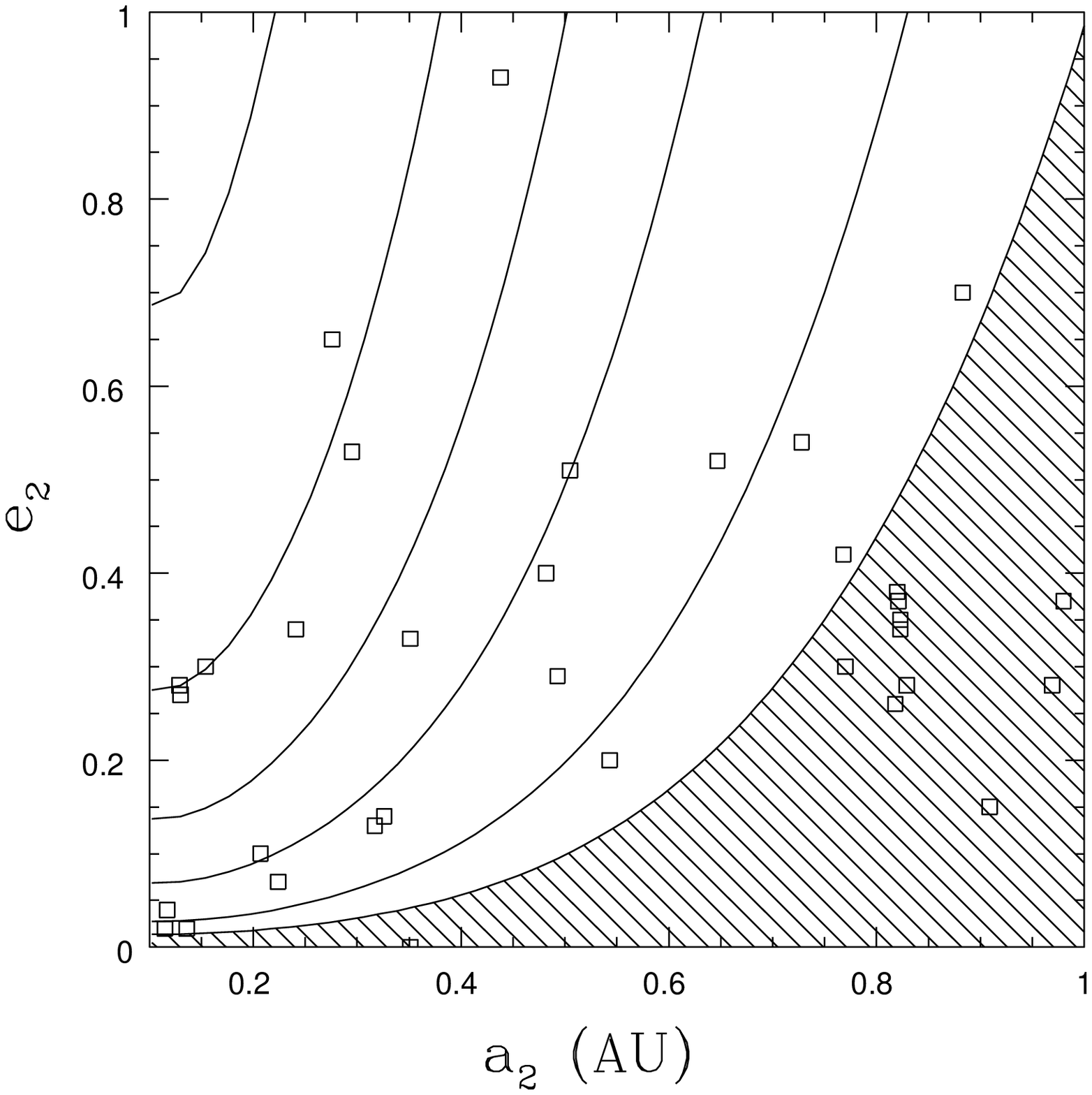} } 
\caption{ For systems that contain a hot Jupiter, a hypothetical
second planet would drive the eccentricity of the inner planet to
larger values.  This plot shows the orbital elements of the second
planet $(a_2,e_2)$ required to excite the eccentricity of the inner
planet to given mean values.  Each curve corresponds to a fixed value
of the mean eccentricity of the hot Jupiter, where $\ebar_1$ = 0.01,
0.02, 0.05, 0.10, 0.20, 0.50 (from bottom to top). The shaded region
delimits the portion of the $a-e$ plane for which additional planets
have a negligible effect on the eccentricity of the hot Jupiter (where
negligible is defined as $\ebar_1 \le 0.01$). For comparison, the open
squares show the orbital elements of the current sample of extrasolar
planets. }
\label{fig:hotjup} 
\end{figure}

[5] For single planet systems with small or zero measured eccentricity
values (e.g., 51 Peg), the small observed $e$ values imply the absence
of large, close planetary companions. For a given system, this line of
reasoning implies a limit on the possible orbital elements of any
additional planets in the system (shown in Fig. \ref{fig:hotjup}). 

The objective of this paper is to outline the effects of secular
interactions in multiple planet systems. We have applied the results
to a small collection of multiple planet systems that have been
detected, and a number of additional applications are expected in the
near future. Although the sample of extrasolar planets ($\sim 150$) is
now large enough to provide some statistical significance, the number
of multiple planet systems is still low ($\sim 20$). Furthermore, the
possible parameter space accessible to multi-planet systems is
enormous, so it is premature to make statistical statements about
multiple planet systems. Nonetheless, future observations will find
greater numbers of systems with multiple giant planets (through
continued radial velocity searches and other methods) and systems
containing smaller, Earth-mass planets (e.g., TPF).

Secular interactions can add to our understanding of these forthcoming
multiple planet systems in a variety of ways. In trying to find
theoretical explanations for the observed orbital elements, one must
take into account the distributions of eccentricities driven by
secular interactions (Fig, \ref{fig:aeplane}).  In systems with known
giant planets, the search for Earths can be guided by studying the
forced eccentricity variations as depicted in Figures
\ref{fig:reshd38529} -- \ref{fig:reshd12661}. Some multiple planet
systems will provide additional (stronger) tests of general relativity
(as in Fig. \ref{fig:genrel}).  In other systems, we can deduce the
presence or absence of additional (undetected) planets --- or at least
constrain their properties --- through examination of the properties
of the detected planets (Fig. \ref{fig:hotjup}). Over longer time
spans, secular interactions combine with tidal circularization and
energy dissipation processes (as outlined in our companion paper Adams
\& Laughlin 2006a).

\bigskip 

This work was supported at U. Michigan (FCA) by the Michigan Center
for Theoretical Physics and by NASA through the Terrestrial Planet
Finder Mission (NNG04G190G) and the Astrophysics Theory Program
(NNG04GK56G0).  This material is based in part upon work supported by
the National Science Foundation CAREER program under Grant No. 0449986
(GL), and was also supported at U. C. Santa Cruz (GL) by NASA through
the Terrestrial Planet Finder Precursor Science Program (NNG04G191G)
and through the Origins of Solar Systems Program (NAG5-13285).

\newpage 


\newpage 

\centerline{\bf Table 1: Secular Time Scales} 
\bigskip 

\begin{center} 
\begin{tabular}{lcccc} 
\hline
\hline 
System & $\tau_1$ (yr) & $\tau_2$ (yr) & $\tau_3$ (yr) & $\tau_4$ (yr)\\ 
\hline 
GJ876    &      81.6 &   53.2 &    4.4 & $-$ \\ 
55Cnc    &   4030000 &   369 &  73.3  & 40.9 \\
UpsAnd   &     4880 &   1924 & 1042 &   $-$ \\
HD217107 &   2865500 &   7198 & $-$ &  $-$ \\
HD160691 &   9431 &  6734 &  2069 & $-$ \\
HD38529  &   1953000  &  10490  & $-$ &  $-$ \\ 
HD190360 &   21055000 &   28360 &  $-$ & $-$ \\
HD11964  &   2118390 & 52260 & $-$ & $-$ \\
HD74156  &    118290 &   9308 & $-$ &  $-$ \\
HD168443 &   13100 & 1837 & $-$ & $-$ \\
HD37124  &   4918 & 2735  & 1047  & $-$ \\
HD73526  &  386 & 55.6 & $-$ & $-$ \\
HD82943  & 551 & 78.9 & $-$ & $-$ \\
HD169830 & 12496 & 3603 & $-$ & $-$ \\
HD202206 & 2130 & 396 & $-$ & $-$ \\
HD12661  & 5501 & 2407 & $-$ & $-$ \\
HD108874 & 6311 & 2261 & $-$ & $-$ \\
HD128311 & 567 & 79 & $-$ & $-$ \\
47UMa    & 5520 & 818 & $-$ & $-$  \\
\hline 
\hline 
\end{tabular} 
\end{center}  

\newpage 
\centerline{\bf Table 2: Eccentricity Variations from Secular Theory} 
\centerline{\bf (systems with 2 planets)}  
\bigskip 

\begin{center} 
\begin{tabular}{lccccccc} 
\hline
\hline 
System & $m_P (m_J)$ & $a$ (AU) & $e_{obs}$ & $\ebar$ & $\sigma_e$ & $e_{min}$ & $e_{max}$ \\ 
\hline
HD217107 b    &   1.369 &   0.074 &   0.130 &   0.130 &   0.0002 &   0.130 &   0.131 \\
HD217107 c    &   1.846 &   3.681 &   0.670 &   0.670 &   0.0000 &   0.670 &   0.670 \\
\hline 
HD38529 b     &   0.783 &   0.129 &   0.275 &   0.274 &   0.0012 &   0.273 &   0.276 \\
HD38529 c     &  12.818 &   3.714 &   0.330 &   0.330 &   0.0000 &   0.330 &   0.330 \\
\hline 
HD190360 c    &   0.062 &   0.128 &   0.010 &   0.117 &   0.0498 &   0.010 &   0.168 \\
HD190360 b    &   1.519 &   3.867 &   0.380 &   0.380 &   0.0000 &   0.380 &   0.380 \\
\hline 
HD11964 b     &   0.116 &   0.230 &   0.150 &   0.148 &   0.0017 &   0.145 &   0.150 \\
HD11964 c     &   0.699 &   3.171 &   0.300 &   0.300 &   0.0001 &   0.300 &   0.300 \\
\hline 
HD74156 b     &   1.858 &   0.294 &   0.635 &   0.617 &   0.0421 &   0.544 &   0.668 \\
HD74156 c     &   6.453 &   3.446 &   0.561 &   0.563 &   0.0038 &   0.558 &   0.569 \\
\hline 
HD168443 b    &   7.784 &   0.295 &   0.529 &   0.538 &   0.0236 &   0.505 &   0.572 \\
HD168443 c    &  17.311 &   2.862 &   0.228 &   0.225 &   0.0082 &   0.212 &   0.236 \\
\hline 
HD73526 b     &   2.039 &   0.652 &   0.390 &   0.382 &   0.0322 &   0.335 &   0.426 \\
HD73526 c     &   2.261 &   1.039 &   0.400 &   0.404 &   0.0217 &   0.373 &   0.434 \\
\hline 
HD82943 b     &   1.856 &   0.746 &   0.380 &   0.295 &   0.1050 &   0.120 &   0.426 \\
HD82943 c     &   1.848 &   1.179 &   0.180 &   0.242 &   0.1056 &   0.051 &   0.369 \\
\hline 
HD169830 b    &   2.881 &   0.812 &   0.310 &   0.357 &   0.0836 &   0.224 &   0.468 \\
HD169830 b    &   4.057 &   3.598 &   0.330 &   0.308 &   0.0323 &   0.259 &   0.353 \\
\hline 
HD202206 b    &  16.786 &   0.811 &   0.435 &   0.416 &   0.0256 &   0.379 &   0.452 \\
HD202206 c    &   2.348 &   2.499 &   0.267 &   0.340 &   0.1382 &   0.095 &   0.508 \\
\hline 
HD12661 b     &   2.339 &   0.822 &   0.330 &   0.305 &   0.0535 &   0.222 &   0.373 \\
HD12661 c     &   1.573 &   2.661 &   0.200 &   0.217 &   0.0618 &   0.123 &   0.299 \\
\hline 
HD108874 b    &   1.386 &   1.042 &   0.130 &   0.298 &   0.1013 &   0.129 &   0.424 \\
HD108874 c    &   0.992 &   2.729 &   0.380 &   0.249 &   0.1075 &   0.060 &   0.380 \\
\hline 
HD128311 b    &   2.475 &   1.077 &   0.210 &   0.265 &   0.1093 &   0.073 &   0.399 \\
HD128311 c    &   3.168 &   1.723 &   0.270 &   0.203 &   0.0890 &   0.043 &   0.311 \\
\hline 
47UMa b      &   2.545 &   2.094 &   0.061 &   0.070 &   0.0133 &   0.050 &   0.088 \\
47UMa c      &   0.764 &   3.735 &   0.100 &   0.074 &   0.0350 &   0.002 &   0.114 \\
\hline  
\hline 
\end{tabular} 
\end{center} 

\newpage 

\centerline{\bf Table 3: Eccentricity Variations from Secular Theory} 
\centerline{\bf (systems with 3 or 4 planets)} 

\bigskip 

\begin{center} 
\begin{tabular}{lccccccc} 
\hline
\hline 
System & $m_P (m_J)$ & $a$ (AU) & $e_{obs}$ & $\ebar$ & $\sigma_e$ & $e_{min}$ & $e_{max}$ \\ 
\hline 
GJ876 d       &   0.019 &   0.021 &   0.000 &   0.013 &   0.0060 &   0.000 &   0.020 \\
GJ876 c       &   0.567 &   0.130 &   0.217 &   0.189 &   0.0209 &   0.158 &   0.217 \\
GJ876 b       &   1.913 &   0.208 &   0.001 &   0.046 &   0.0221 &   0.001 &   0.072 \\
\hline
55Cnc e       &   0.045 &   0.038 &   0.174 &   0.185 &   0.0631 &   0.043 &   0.301 \\
55Cnc b       &   0.827 &   0.115 &   0.061 &   0.128 &   0.0608 &   0.009 &   0.203 \\
55Cnc c       &   0.173 &   0.241 &   0.493 &   0.430 &   0.0552 &   0.347 &   0.506 \\
55Cnc d       &   3.746 &   5.526 &   0.251 &   0.251 &   0.0000 &   0.251 &   0.251 \\
\hline
UpsAnd b      &   0.686 &   0.059 &   0.011 &   0.016 &   0.0062 &   0.002 &   0.027 \\
UpsAnd c      &   1.913 &   0.842 &   0.268 &   0.177 &   0.0812 &   0.025 &   0.274 \\
UpsAnd d      &   3.760 &   2.527 &   0.270 &   0.288 &   0.0135 &   0.268 &   0.306 \\
\hline
HD160691 d    &   0.046 &   0.090 &   0.000 &   0.035 &   0.0136 &   0.000 &   0.051 \\
HD160691 b    &   1.669 &   1.500 &   0.200 &   0.489 &   0.1931 &   0.156 &   0.727 \\
HD160691 c    &   3.135 &   4.170 &   0.570 &   0.496 &   0.0576 &   0.411 &   0.574 \\
\hline
HD37124 b     &   0.710 &   0.546 &   0.071 &   0.130 &   0.0487 &   0.026 &   0.203 \\
HD37124 c     &   0.695 &   1.743 &   0.170 &   0.115 &   0.0527 &   0.009 &   0.187 \\
HD37124 d     &   0.739 &   3.078 &   0.100 &   0.108 &   0.0395 &   0.024 &   0.179 \\
\hline
\hline 
\end{tabular} 
\end{center}

\newpage 

\newpage 

\newpage 

\newpage 

\newpage 

\newpage 

\newpage 

\newpage 

\newpage 

\newpage 

\newpage 


\begin{thebibliography} 

\bibitem[abst]{abst}
Abramowitz, M., \& Stegun, I. A. 1970, Handbook of Mathematical Functions
(New York: Dover) 

\bibitem[al]{al} 
Adams, F. C., \& Laughlin, G. 2003, Icarus, 163, 290

\bibitem[companion]{compansion}
Adams, F. C., \& Laughlin, G. 2006a, submitted to ApJ 

\bibitem[algrav]{algrav} 
Adams, F. C., \& Laughlin, G. 2006b, submitted to Grav. Res. Found.
Essays on Gravitation

\bibitem[br]{br} 
Barnes, R., \& Raymond, S. N. 2004, ApJ, 617, 569 

\bibitem[bw]{bw} 
Brouwer, D.  \& van Woerkom, A.J.J.  1950, Astron. P. Amer. Ephem.,
13, 81

\bibitem[b99]{b99} 
Butler, R. P., Marcy, G. W., Fischer, D. A., Brown, T. M., Contos, A. R., 
Korzennik, S. G., Nisenson, P., \& Noyes, R. W. 1999, ApJ, 526, 916 

\bibitem[david]{david} 
David, E. M., Quintana, E. V., Fatuzzo, M., \& Adams, F. C. 2003,
PASP, 115, 825

\bibitem[erdi]{erdi} 
{\'E}rdi, B., Dvorak, R., S{\'a}dnor, Zs., Pilat-Lohinger, E., \& 
Funk. B. 2004, MNRAS, 351, 1043 


\bibitem[ford1]{ford1}
Ford, E. B., Kozinsky, B., \& Rasio, F. A. 2000a, ApJ, 535, 385 

\bibitem[ford2]{ford2}
Ford, E. B., Joshi, K. J., Rasio, F. A., \& Zbarsky, B. 
2000b, ApJ, 528, 336 

\bibitem[gold]{goldreich} 
Goldreich, P., \& Sari, R. 2003, ApJ, 585, 1024 




\bibitem[htt]{htt}
Holman, M., Touma, J., \& Tremaine, S. 1997, Nature, 386, 254 


\bibitem[]{jsc01} 
Jones, B. W., Sleep, P. N., Chambers, J. E. 2001, A\&A, 366, 254  

\bibitem[kwr]{kwr} 
Kasting, J. F., Whitmire, D. P., \& Reynolds, R. T. 1993, Icarus, 101, 108

\bibitem[lask88]{lask88}
Laskar, J. 1988, A \& A, 198, 341  

\bibitem[lp]{lp}
Lee, M. H., \& Peale, S. J. 2003, ApJ, 592, 1201 

\bibitem[lo]{lo} 
Lubow, S. H., \& Ogilvie, G. I. 2001, ApJ, 560, 997   

\bibitem[lunine]{lunine} 
Lunine, J. I. 2005, Astrobiology: A Multidisciplinary Approach 
(San Francisco: Pearson Educational Press) 

\bibitem[mb]{mb} 
Marcy, G., \& Butler, P. R. 1996, ApJ, 464, L147 

\bibitem[ml02]{ml02} 
Mardling, R. A., \& Lin, D.N.C. 2002, ApJ, 573, 829 

\bibitem[ml04]{ml04} 
Mardling, R. A., \& Lin, D.N.C. 2004, ApJ, 614, 955 

\bibitem[mq]{mq} 
Mayor, M., \& Queloz, D. 1995, Nature, 378, 355 

\bibitem[]{mt03} 
Menou, K., \& Tabachnik, S. 2003, ApJ, 583, 473 

\bibitem[mm04]{mm04} 
Michtchenko, T. A., \& Malhotra, R. 2004, Icarus, 168, 237

\bibitem[ma]{ma} 
Moorhead, A. V., \& Adams, F. C. 2005, Icarus, 178, 517 

\bibitem[md]{md99} 
Murray, C. D., \& Dermott, S. F. 1999, Solar System Dynamics
(Cambridge: Cambridge Univ. Press) (MD99) 

\bibitem[naga]{naga} 
Nagasawa, M., Lin, D.N.C., \& Ida, S. 2003, ApJ, 586, 1374 

\bibitem[]{nmc02} 
Noble, M., Musielak, Z. E., Cuntz, M. 2002, ApJ, 572, 1024 

\bibitem[oglu]{oglu} 
Ogilvie, G. I., \& Lubow, S. H.  2003, ApJ, 587, 398  

\bibitem[rivera]{rivera} 
Rivera, E. J. et al. 2005, ApJ, 634, 625 

\bibitem[takeda]{takeda} 
Takeda, G., \& Rasio, F. A. 2005, ApJ, 627, 1001 

\bibitem[wug]{wug} 
Wu, Y., \& Goldreich, P., 2002, ApJ, 564, 1024 

\end{thebibliography}
\end{document}